# Atomistic Simulations of Cation Distribution and Defect Effects on the Performance of Substituted Ferrites

Jiahao Li[1], Kusma Kumari Cheepurupalli[2], Niall J. English[3], Sateesh Bandaru[1*], Xuefeng Zhang[*1]

[1]Key Laboratory of Novel Materials for Sensor of Zhejiang Province College of Materials and Environmental Engineering, Hangzhou Dianzi University, Hangzhou 310018, China.

[2]Gayatri Vidya Parishad College of Engineering (Autonomous), Visakhapatnam, Andhra Pradesh, India-530 048

[3]School of Chemical and Bioprocess Engineering, University College Dublin, Belfield, Dublin 4, Ireland

*corresponding authors: e-mail: sbandaru@hdu.edu.cn, zhang@hdu.edu.cn

**Abstract**

Nanoscale spinel-structured ferrites substituted with multivalent cations have garnered significant attention due to their tunable functional properties, making them highly relevant for high-frequency electronics, energy conversion, and electromagnetic shielding applications. In this study, we investigate Mn-Zn ferrites with a nominal composition of $Mn_{0.5}Zn_{0.5}Fe_2O_4$ (MZF), substituted with tetravalent ($Si^{4+}$), trivalent ($Co^{3+}$), and divalent ($Ca^{2+}$, $Mg^{2+}$, $Sn^{2+}$) ions. A comprehensive analysis is conducted to understand how these substitutions at specific tetrahedral and octahedral crystallographic sites influence the structural stability, electronic band structure, magnetic anisotropy, and electrical conductivity of the spinel lattice. Density functional theory (DFT), combined with Boltzmann transport theory, is employed to investigate the thermoelectric and phonon transport properties of pristine and doped MZF systems. Formation energy calculations reveal that substitutions with $Si^{4+}$, $Ca^{2+}$, and $Mg^{2+}$ enhance the thermodynamic stability of MZF, whereas $Co^{3+}$ and $Sn^{2+}$ introduce slightly higher formation energies, indicating relatively lower stability. Electronic structure analyses confirm that all substituted variants maintain a finite band gap, preserving their semiconducting nature. Magnetic anisotropy energy (MAE) calculations show that ferrites with mixed octahedral tetrahedral substitutions exhibit a narrower MAE distribution, suggesting more uniform magnetic anisotropy. This characteristic is associated with lower coercivity ($H_c$) and reduced hysteresis losses key attributes for soft magnetic materials operating at high frequencies. In contrast, oxygen-deficient (Ov) MZF structures display a little broader MAE distribution, indicative of increased magnetic disorder, which may contribute to elevated eddy current losses under alternating magnetic fields.

Thermoelectric property analysis at 300 K demonstrates that multivalent ion doping at either crystallographic site leads to a reduction in electrical conductivity ($\sigma$), while concurrently enhancing the Seebeck coefficient (S). This inverse correlation highlights a doping-induced trade-off, likely driven by increased carrier scattering at defect sites and modifications to the electronic density of states near the Fermi level. Overall, the findings underscore the critical role of substitutional chemistry, crystallographic site specificity, and defect engineering in modulating the magnetic, thermal, and thermoelectric properties of MnZn ferrites. These insights are pivotal for the design of next-generation functional materials for high-efficiency magnetic and energy conversion devices.

**Introduction**

Significant progress has been made in materials science and nanotechnology in recent years, resulting in ground-breaking discoveries in a number of domains. The domains such as, medicine, energy, electronics, and environmental research have all been greatly impacted by developments in soft/hard ferrite materials.[1-8] These advancements have unlocked new possibilities for designing multifunctional, sustainable, and highly efficient materials. Owing to their exceptional properties, ferrites particularly hybrid doped structures exhibit immense potential for diverse applications [9-13]. Among them, manganese-zinc ferrite, with its face-centered cubic spinel structure (space group $Fd\bar{3}m$), has become indispensable in high-frequency technologies.[14-15] MnZn-ferrites represent the gold standard for soft magnetic materials in advanced power electronics, driven by their exceptional high-frequency performance. These materials exhibit ultra-high initial permeability ($\mu_i$ > 15,000) and high saturation magnetization (~500 mT), with minimal thermal/ frequency-dependent degradation properties critical for MHz-range applications such as high-efficiency transformers, miniaturized inductors, and EMI suppression devices. Their magnetic performance is governed by cation distribution tetrahedral (Td) vs. octahedral (Oh) site occupancy, sintering protocols, stoichiometric control, and synthesis methods, which collectively tune grain boundary resistivity, domain wall motion, and core loss mechanisms ($P_{cv} = P_h + P_e + P_r$). As the backbone of modern power systems, MnZn-ferrites enable compact, energy-dense designs for switching power supplies, broadband signal isolation, and electromagnetic compatibility in next-generation grid/telecom infrastructure. As 5G networks, electric vehicles, and green energy solutions revolutionize our technological landscape, MnZn ferrites face their greatest challenge yet: conquering the MHz frequency frontier while slashing energy losses.[16-17] Power loss (Pcv), a critical property of MnZn ferrites, comprises three components: hysteresis loss ($P_h$), eddy-current loss ($P_e$), and residual loss ($P_r$).[18–20] To optimize these losses and enhance magnetic performance, ion doping has been a conventional strategy to tailor the microstructure and electrical properties of MnZn ferrites. To address these performance constraints, researchers have extensively explored the influence of ion doping on manganese-zinc ferrites, analyzing both its modification effects and mechanistic basis. For instance, doping with 0.3 wt% $Co^{2+}$ enables its substitution at the octahedral (B) sites of $Fe^{3+}$, reducing the magnetocrystalline anisotropy constant ($K_1$) and thereby decreasing hysteresis losses by 15-20%.[21] Doping with 0.5 mol% $Ti^{4+}$ inhibits the generation of $Fe^{2+}$ ions and minimizes oxygen vacancies near grain boundaries, leading to a 40% reduction in eddy current losses relative to undoped materials.[22] The incorporation of rare-earth elements (e.g., $La^{3+}$, $Yb^{3+}$) suppresses grain growth through grain boundary pinning, refining the microstructure to sub-micron dimensions (<1 μm). This structural modification significantly improves thermal stability, reducing the permeability attenuation rate from 12% to 4% at elevated temperatures (150°C).[23] Extensive studies have been conducted on cation distribution ordering behavior in spinel ferrites. Extensive investigations employing thermodynamic modeling and first-principles calculations have elucidated the cation site preference in spinel ferrites and its impact on magnetic properties. The analysis reveals distinct occupancy behaviors: $Mn^{2+}$ exhibits strong octahedral site preference, while $Zn^{2+}$ demonstrates pronounced tetrahedral site selectivity. This differential cation distribution fundamentally governs the material's magnetic characteristics, with $Mn^{2+}$ occupancy enhancing magnetic anisotropy through spin-orbit coupling at octahedral sites, and $Zn^{2+}$ incorporation in tetrahedral sites directly influencing saturation magnetization by modifying the A-B super exchange interaction pathways.[24] MnZn-ferrite nanoparticles, diameter: 20-40 nm) were synthesized via an optimized Pechini sol-gel method. Density functional theory (DFT) calculations were employed to elucidate the mechanisms underlying their enhanced permittivity and permeability. The analysis revealed that the improved dielectric properties stem from a reduction in surface oxygen vacancies and enhanced electron localization at grain boundaries. These structural modifications suppress polarization losses, resulting in a 20% reduction in dielectric loss across the 2-18 GHz frequency range.[25] Despite extensive research on specific characteristics, this class of ferrites has received comparatively less attention for advancing key properties such as high-frequency efficiency, strong saturation magnetization, and the development of Fe-based nanocrystalline materials. A comprehensive analysis of these structures is therefore critical to unlock their potential for diverse applications, particularly in optimizing soft ferrites for next-generation technologies.

In the conventional spinel structure, $Zn^{2+}$ ions typically occupy the tetrahedral (A) sites, while $Mn^{2+}$ and $Fe^{3+}$ ions predominantly reside at the octahedral (B) sites. In this study, we investigate the site-specific substitution of foreign elements at both tetrahedral (Td) and octahedral (Oh) positions originally occupied by $Fe^{3+}$ in MZF. This targeted doping leads to significant changes in the structural, magnetic, and electrical properties of the ferrite lattice, which are systematically analyzed and discussed. Additionally, we explore the role of oxygen vacancies in influencing the performance of ferrites by predicting defect formation energies and assessing their effects on conductivity and magnetic behavior. Despite the widespread use of density functional theory (DFT+U) in such analyses, it faces limitations in accurately capturing magnetic anisotropy energy (MAE) in soft ferrites an essential factor for tuning magnetic permeability and minimizing energy losses. Therefore, this work employs first-principles simulations to study the electrical, magnetocrystalline anisotropy, and dielectric properties of both pure and doped MZF's, incorporating tetravalent ($Si^{4+}$, $Sn^{4+}$), trivalent ($Co^{3+}$), and divalent ($Ca^{2+}$, $Mg^{2+}$) ions.

**Computational Methods**

First-principles calculations were performed using the Vienna Ab Initio Simulation Package (VASP).[26,27] The projector-augmented wave (PAW) method [28] was employed to describe electron-ion interactions, with the following valence electron configurations: Fe ($3d^74s^1$), Mn ($3d^64s^1$), P ($3s^23p^3$), Zn ($3d^{10}4s^2$), and O ($2s^22p^4$), as implemented by Kresse and Joubert. To account for strong electron correlations in the 3d orbitals of Fe, the spin-polarized GGA+U approach was applied within Dudarev's formalism.[29-30] This method introduces an effective Hubbard correction ($U_{eff}$) for the Fe 3d states, incorporating an on-site Coulomb repulsion term (U) into the DFT Hamiltonian to improve the description of localized electrons and electronic bandgap predictions. The DFT+U functional is expressed as (Eq1):

$$E_{DFT+U} = E_{LSDA} + \frac{(U-J)}{2} \sum_\sigma \left[ \left( \sum_{m_1} n^\sigma_{m_1,m_1} \right) - \left( \sum_{m_1,m_2} \hat{n}^\sigma_{m_1,m_2} \hat{n}^\sigma_{m_2,m_1} \right) \right] \ldots \text{Eq1}$$

$U_{eff}$ was set to 4.3 and 3.5 eV for Fe and Mn atoms and for Zn 6.0 eV, respectively. The number of VASP plane-waves was controlled by the cut-off energy, which we set to 550 eV, having tested for cut-off convergence. Brillouin-zone integrations were performed during geometry relaxation using k-point grids with a $4 \times 4 \times 4$ for MZF and the doped structures MZF (M@MZF). A van der

Waals correction was examined for vdW functionals using the Grimme method in conjunction with Becke-Jonson damping.[31] Structure optimization was performed until the convergence criteria of energy and force were achieved, which were $10^{-6}$ eV and 0.01 eV/Å respectively. Doping formation energies were calculated using following formulas (Eq2 and Eq3) [32-33]: The substitution energy of M (Co, Ca, Mg, Sn, Si) in MZF i.e., the change in the formation energy of M-substituted at octahedral/tetrahedral site is M@MZF with respect to MZF is

$$E_{form} = E_{M@MZF} - E_{MZF} - \mu E_M + \mu E_{Fe} \ldots \ldots \quad Eq2$$

$$E_{form} = E_{M@Ov-MZF} - E_{Ov-MZF} - \mu E_M + \mu E_{Fe} \ldots Eq3$$

Where, $E_{M@MZF}$, $E_{MZF}$ are the energies of the Metal and non-metal doped MZF; and $E_{MFZ}$ is $Mn_{0.5}Zn_{0.5}\text{-}Fe_2O_4$, respectively. $\mu E_{Fe}$ and $\mu E_M$ respectively represent the energy of isolated Fe atom and impurity metal/non-metal atoms respective stable crystal forms. $E_{M@Ov-MZF}$, $E_{Ov-MZF}$ are the energies presence of oxygen vacancy doped MZF systems and O-vacant MZF structure. Furthermore, the chemical potential-dependent electrical conductivity (Eq4) and Seebeck coefficient (Eq5) have been calculated by integrating the transport distribution function as follows:

$$\sigma_{\alpha\beta}(\alpha,\mu) = \frac{1}{\Omega} \int \sigma_{\sigma\beta}(\varepsilon) \left[ -\frac{\partial f_0(T,\varepsilon,\mu)}{\partial \varepsilon} \right] d\varepsilon \quad \ldots \ldots Eq4$$

$$S_{\alpha\beta}(T,\mu) = \frac{1}{eT\Omega\sigma_{\alpha\beta}(T,\mu)} \int \sigma_{\sigma\beta}(\varepsilon)(\varepsilon-\mu) \left[ -\frac{\partial f_0(T,\varepsilon,\mu)}{\partial \varepsilon} \right] d\varepsilon \quad \ldots Eq5$$

The BoltzTraP code[34] has been employed to implement the above theory, enabling the analysis of thermoelectric properties such as electrical conductivity and Seebeck coefficient. Seebeck coefficients (thermo-power) can be used to study thermoelectric potentials. Materials with high Seebeck coefficients have good ability to push electrons from hot regions to cold regions. Based on the rigid band approximation and constant scattering time approximation, we calculated the thermoelectric properties of all these ferrites. We calculate transport distribution tensors $\sigma_{\alpha\beta}(\varepsilon)$ by interpolating the electronic band structure.

Further, the magnetocrystalline anisotropy energy (MAE) is determined via DFT+U with spin-orbit coupling (SOC) included, defined as: $E_{MAE} = E_{[001]} - E_{[100]}$, where $E_{[001]}$ and $E_{[100]}$, represent the total energies of the doped MnZn-ferrite system with magnetization oriented along the [001] and [100] directions, respectively. The total energies for in-plane [100] and out-of-plane [001] magnetization directions are calculated self-consistently using the force theorem.[35-36] For these calculations, the cutoff energy was set to 600 eV, and the convergence criterion for energy was tightened to $10^{-8}$ eV to ensure high precision in the MAE determination.

The interaction of photons with electrons in a system gives rise to its optical properties, which can be described using time-dependent perturbation theory applied to the ground-state electronic states. The imaginary part of the dielectric function, $\varepsilon_2(\omega)$ can be calculated in a manner analogous to Fermi's Golden Rule for time-dependent perturbations, accounting for real transitions between occupied and unoccupied states. This allows for the determination of the dielectric properties of both pure and doped MZF systems. By applying the appropriate selection rules,[37] the momentum matrix elements between occupied and unoccupied wave functions can be used to compute $\varepsilon_2(\omega)$. The real part of the dielectric function $\varepsilon_1(\omega)$, along with the absorption coefficient $\alpha(\omega)$, can then be derived using the Kramers-Kronig relations (Eq6).

$$\alpha(\omega) = \sqrt{2}\omega\left[\sqrt{\varepsilon_1^2(\omega) + \varepsilon_2^2(\omega)} - \varepsilon_1(\omega)\right]^{\frac{1}{2}} \quad \ldots\ldots Eq6$$

**Results & Discussion**

In this study, we explore the impact of doping $Mn_{0.5}Zn_{0.5}Fe_2O_4$ with foreign elements (M=Ca, Co, Mg, Si, Sn) through site-specific substitutions. The parent spinel structure features $Zn^{2+}$ ions occupying tetrahedral (A) sites, while $Mn^{2+}$ and $Fe^{3+}$ ions predominantly reside at octahedral (B) sites. Our focus lies on systematically replacing $Fe^{3+}$ ions at both tetrahedral and octahedral sites with the dopant elements (M), forming the modified structure $Mn_{0.5}Zn_{0.5}M_xFe_{2-x}O_4$ (M@MZF). This targeted substitution induces structural, magnetic, and electrical alterations in the ferrite lattice, which are comprehensively analyzed. The MZF unit cell comprises 56 atoms: 32 oxygen anions and 24 cations. Of the cations, 8 occupy tetrahedral (A) sites (shared equally by $Mn^{2+}$ and $Zn^{2+}$), while the remaining 16 occupy octahedral (B) sites (dominated by $Fe^{3+}$ and $Mn^{2+}$). By substituting $Fe^{3+}$ ions at both A and B sites with foreign dopants, we probe the resulting modifications in the material's properties. These site-engineered changes are investigated to understand their influence on lattice dynamics, magnetic behavior, and electrical conductivity, offering insights for tailored ferrite design. MZF was successfully synthesized using three established methods: (i) conventional solid-state reaction[38], (ii) controlled co-precipitation,[39] and (iii) hydrothermal synthesis.[40] X-ray diffraction (XRD) analysis confirmed the formation of a single-phase spinel structure, with all diffraction peaks indexed to the cubic spinel phase belonging to the $Fd\bar{3}m$ space group. Structural characterization, as depicted in Fig 1(a), confirms the phase purity and high crystallinity of the synthesized MZF. Electronic band structure analyses (Figs 1(b) and 1(c)) demonstrate semiconducting behavior in the material, with both the valence band maximum (VB) and conduction band minimum (CB) situated at the Γ point, signifying a direct bandgap. Further insights from the total and partial density of states (DOS, Fig 1(b)) reveal distinct orbital contributions: the conduction band arises primarily from hybridization of Mn 3d and O 2p orbitals, while the valence band is dominated by O 2p states. This strong Mn-O orbital overlap underscores the covalent interaction governing charge transport in the material. Magnetic property calculations for the MZF unit cell predict a total magnetic moment of 92.0 µB, reflecting the collective alignment of spins from transition-metal cations ($Mn^{2+}$, $Fe^{3+}$) at octahedral sites. These results collectively elucidate the interplay between atomic-scale substitutions, electronic structure, and macroscopic magnetic behavior in the doped ferrite system. The calculated average magnetic moments of 4.32 µB per Fe atom and 3.86 µB per Mn atom underscore the dominant role of transition-metal cations in driving the magnetism of the spinel lattice. In this work, we systematically investigate the substitution of foreign elements (Ca, Co, Mg, Si, Sn) into both tetrahedral (Td) and octahedral (Oh) sites of MZF, with a focus on site-specific doping strategies. Additionally, we explore the interplay between dopant incorporation and oxygen vacancies within the lattice, probing how these defects influence the crystallographic site preferences and electronic environment of the substituted elements. This dual approach targeted substitution at Td/Oh sites and defect engineering through oxygen vacancies enables a comprehensive understanding of how atomic-scale modifications alter the magnetic, electronic, and structural properties of doped MZF ferrites.

**Formation energies & Lattice distortion**

To systematically investigate dopant effects on MZF, we employed density functional theory (DFT+U) calculations by substituting a single $Fe^{3+}$ ion at both tetrahedral (Td, A-site; Fig. S1) and octahedral (Oh, B-site; Fig. S2) positions with $Ca^{2+}$, $Co^{3+}$, $Si^{4+}$, $Mg^{2+}$, or $Sn^{4+}$. This methodology was further extended to oxygen-deficient systems (M@Ov-MZF; Fig. S3), incorporating both dopants and engineered oxygen vacancies. Fully relaxed structures revealed notable dopant-induced lattice distortions relative to pristine MZF. The observed changes in lattice parameters and cell volumes (Fig. S1) were primarily governed by the dopant ionic radius compared to $Fe^{3+}$

(0.645 Å in Td and 0.785 Å in Oh coordination). Minimal lattice distortion ($\Delta a < 0.5\%$) was observed for $Co^{3+}$ substitution at the Oh site (ionic radius 0.72 Å), attributed to its close match with $Fe^{3+}$. In contrast, significant lattice expansion ($\Delta a \approx 1.8–3.1\%$) was induced by larger dopants such as $Ca^{2+}$ (Oh site, 1.00 Å) and $Sn^{4+}$ (Td site, 0.55 Å). Conversely, $Si^{4+}$ substitution (Td site, 0.40 Å) led to a moderate lattice contraction ($\Delta a \approx -0.7\%$) due to its smaller size, while $Mg^{2+}$ (Oh site, 0.72 Å) caused a slight expansion. DFT-calculated formation energies (Figs. 1d, 1e) revealed distinct thermodynamic trends. Ca-, Si-, and Mg-doped configurations exhibited negative formation energies ($< -1.0$ eV), indicating thermodynamically favorable incorporation and intrinsic stability. In contrast, Sn-doped systems showed significantly positive formation energies (~5.5–6.0 eV), suggesting metastable behavior under ambient conditions. This preferential Td-site substitution correlates with enhanced lattice relaxation capabilities in vacancy-rich environments, suggesting a defect-mediated stabilization mechanism.

Metal ion doping introduces lattice distortions that can significantly influence the physical, chemical, and electronic properties of materials, which can be exploited for a range of applications such as catalysis, electronics, and optics. When foreign elements such as Ca, Co, Mg, Si, and Sn are doped into lattices, they can substitute for the native metal ions (Tetrahedral/Octahedral $Fe^{3+}$) in the host lattice, leading to lattice distortions and modifications in the material's properties. Here's how these substitutions typically occur and their effects. $Ca^{2+}$ ions usually substitute for cations like $Fe^{3+}$ in the tetrahedral (A) or octahedral (B) sites. Due to its relatively large ionic radius (~1.00 Å for CN=6), Ca doping can introduce strain and lattice expansion. $Co^{2+}$ prefers the octahedral (B) site due to its strong crystal field stabilization energy. It substitutes for $Fe^{3+}$, causing local distortions due to its different ionic radius (~0.745 Å for high-spin $Co^{2+}$, CN=6) and John-Teller effects. $Mg^{2+}$ (~0.72 Å, CN=6) typically replaces $Fe^{3+}$ in the tetrahedral or octahedral sites, respectively, due to slight lattice distortion. $Si^{4+}$ is a small ion (~0.40 Å, CN=4) and usually occupies tetrahedral sites, replacing $Fe^{3+}$. The high charge mismatch ($M^{4+}$ vs. $M^{2+/3+}$) may require charge compensation mechanisms (e.g., cation vacancies or $Fe^{2+}$ formation), further distorting the lattice. $Sn^{4+}$ (~0.69 Å, CN=6) substitutes for $Fe^{3+}$ in octahedral sites and also tetrahedral sites. Its larger size and higher charge can induce strain and alter magnetic exchange interactions. Foreign ion doping in MZF introduces lattice strain and property modifications by substituting host ions, with effects dependent on the dopant's ionic radius, charge, and site preference.

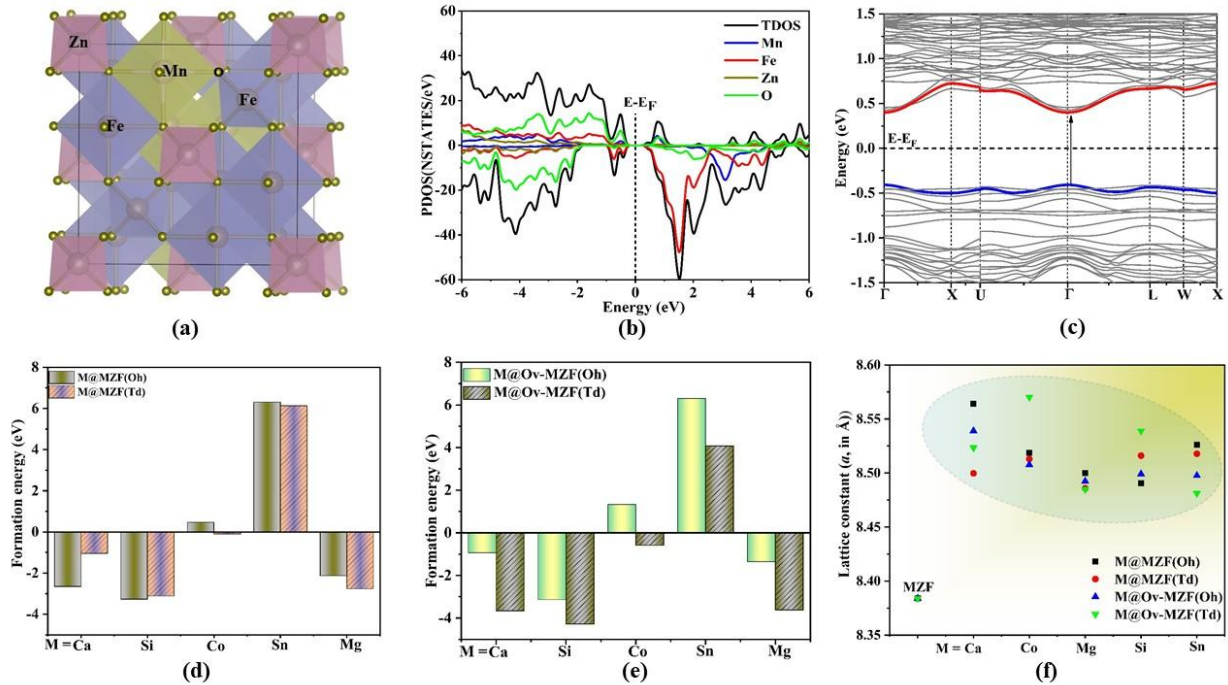

**Fig 1.** (a) Fully relaxed structure of MZF. (b) and (c) depict projected density of states (PDOS) and the band structure of the MZF structure. (d)-(e) and (f) show the formation energies and Lattice distortion in MZF and the effect of foreign element doping at tetrahedral and octahedral sites in both pristine MZF and oxygen- vacant MZF (Ov-MZF) respectively.

Systematic analysis (Fig S4) confirms these trends follow Shannon-Prewitt radii for tetrahedral coordination: $Fe^{3+}$ (0.49 Å) < $Sn^{4+}$ (0.55 Å) < $Co^{2+}$ (0.72 Å) < $Ca^{2+}$ (1.00 Å). The monotonic increase in *c/a* ratio (Table S1) with dopant size underscores cationic steric effects as the primary driver of lattice distortion. This establishes ionic radius mismatch as a predictive metric for strain engineering in spinel ferrites. Controlled doping is key to engineering ferrites for specific applications.

**Effect of Cation Distribution and Defect sites on the Electronic Structure**

To gain deeper insight into the impact of dopants on the electronic properties of $Mn_{0.5}Zn_{0.5}Fe_2O_4$ (MZF), we conducted comprehensive electronic structure calculations for all doped systems, encompassing both MZF and oxygen-deficient (Ov-MZF) configurations. These analyses reveal how different dopants modulate the electronic characteristics of the host lattice, offering a detailed understanding of dopant-induced electronic modifications. The total and partial density of states (DOS), along with the corresponding band structures, are shown in Figs 2 and 3 for Ca-, Si-, and Mg-doped MZF, with dopant substitutions at both octahedral and tetrahedral sites selected based on their relatively favorable formation energies.

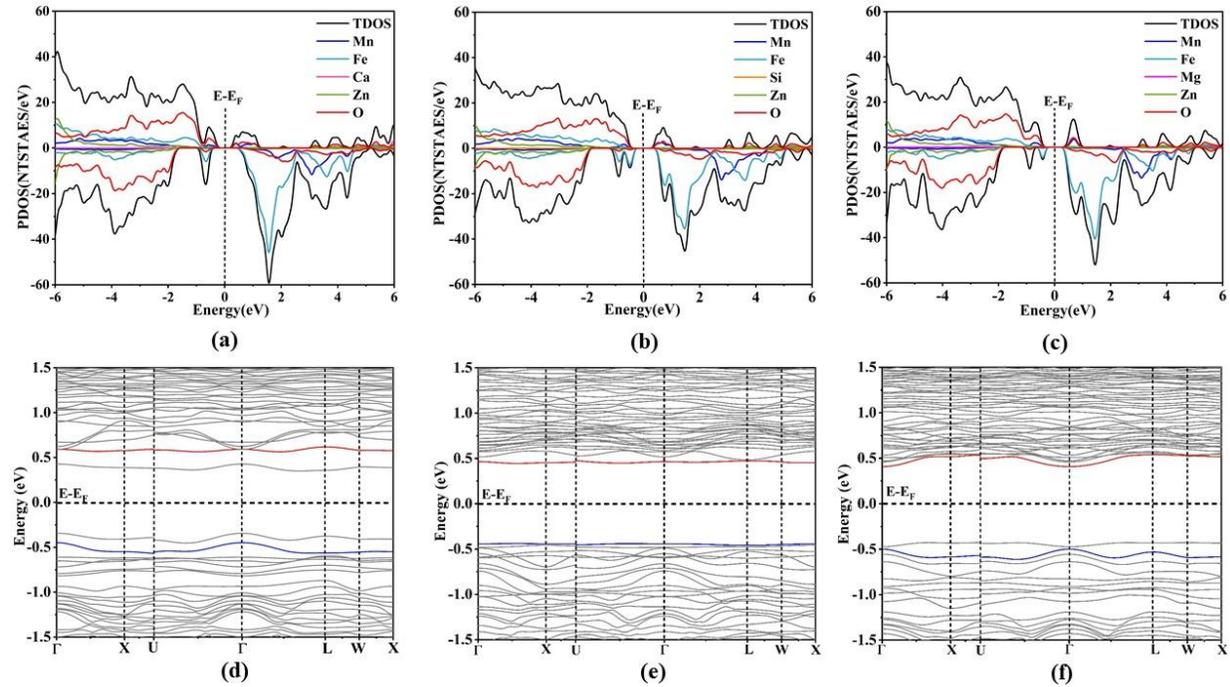

**Fig 2.** (a)-(f) present the total and projected densities of states (PDOS) and band structures for M-doped MZF (where M = Ca, Si, and Mg), calculated using GGA+U. In these systems, the foreign elements are substituted at the octahedral sites of the MnZn-ferrite (MZF) lattice.

The electronic structure results for Co- and Sn-doped systems at both sites are provided in the Supporting Information (Figs. S5 and S6). These results reveal key electronic modifications induced by doping, such as shifts in Fermi level, emergence of impurity states, and alterations in bandgap behavior. These data, calculated for M-doped systems $Mn_{0.5}Zn_{0.5}M_xFe_{2-x}O_4$ at both octahedral and tetrahedral sites, enable comparative assessment across dopants and structural configurations. These detailed electronic structure analyses are essential for understanding the mechanistic effects of doping on the electronic properties of both MZF and Ov-MZF systems. The total and partial density of states (DOS) reveal that the valence bands are primarily formed by the O 2p orbitals, while the conduction bands mainly originate from Mn 3d and O 2p orbitals.

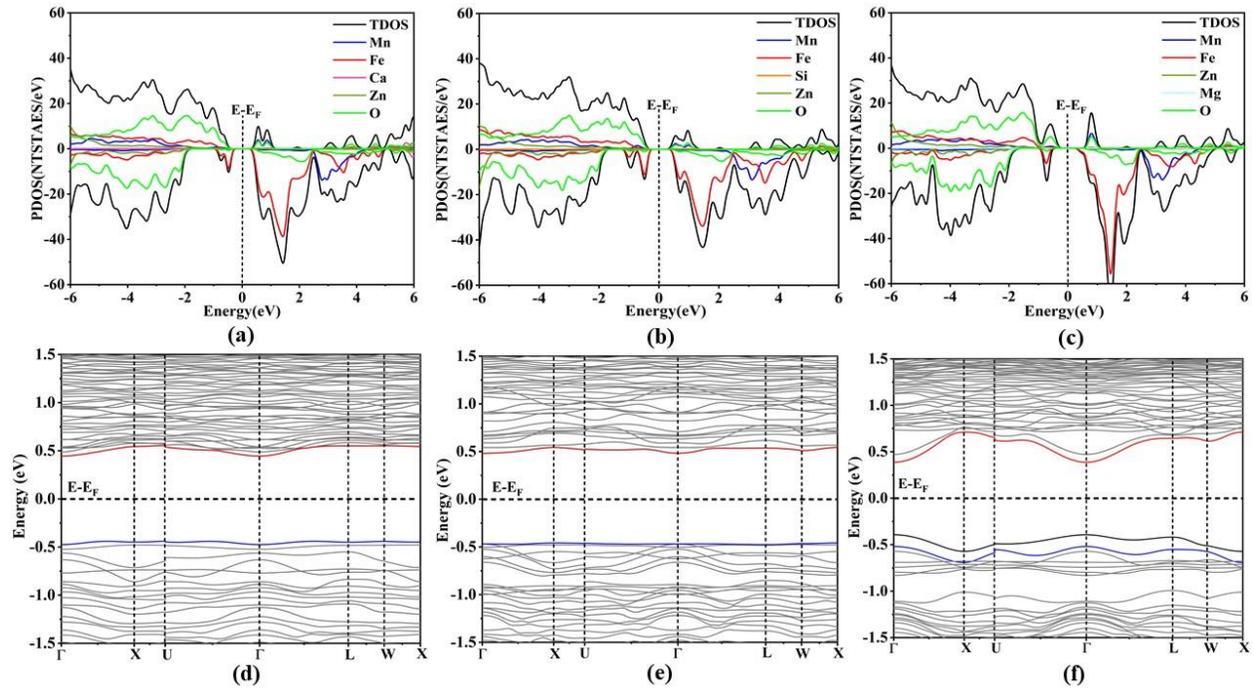

**Fig 3.** (a)-(f) show the total and projected densities of states (PDOS) and band structures for M-doped M@MZF (where M = Ca, Si, and Mg), calculated using GGA+U. In these systems, the foreign elements are doped at the tetrahedral sites of the MnZn-ferrite (MZF) lattice.

Analysis of the partial DOS and band structures confirms that all doped MZF systems retain their semiconducting nature. The computed band gap values for these doped structures, provided in Table S1, show only slight variations (ranging from 0.1 to 0.2 eV). These minor changes are attributed to hybridization and orbital interactions between the dopant atoms (Ca 3d, Mg 2p, and Si 3p) and the host lattice. Doping primarily affects the valence band structure, with Si and Ca notably introducing mid-gap states (Fig 2, 2(b) and (f)) due to its p-orbital interactions. These mid-gap states are clearly visible within the band gap region of the corresponding band structure plots. DFT calculations reveal site-dependent electronic perturbations, showing that tetrahedral-site incorporation of Ca, Mg, Si, Sn, and Co dopants induces distinct modifications to MZF's band structure and density of states. The computed band gaps are systematically lower (by ~0.1-0.3 eV) compared to octahedral-site. Complete bandgap data are tabulated in Supporting Tables S1 and S2. If we compare the PDOS of these structures, reduced orbital overlap at tetrahedral sites, and altered cation-anion bond angles affecting p-d hybridization. Si and Sn show the most pronounced effects due to their strong sp hybridization. This site-dependent bandgap engineering suggests tetrahedral doping as a viable strategy for tuning MZF's optoelectronic properties in high-frequency applications.

Further investigations were conducted to examine the impact of oxygen vacancies in MZF, along with the substitution of metal/non-metal atoms at both tetrahedral and octahedral sites. The calculated oxygen vacancy (Ov) formation energy in MnZn-ferrite (MZF) was found to be thermodynamically favorable, with a value of -0.81 eV, indicative of spontaneous vacancy formation under the studied conditions. Notably, substitutions at oxygen vacancy sites with Ca, Si, and Mg further reduced the formation energy, as demonstrated in Fig 1(e), suggesting enhanced stability for these doped systems (M@Ov-MZF). Complementing these findings, the electronic structure analysis including the computed density of states (DOS) and band structures (Figs S7 and S8) revealed defect-induced modifications, such as mid-gap states and altered carrier mobility, which correlate with the observed trends in vacancy stability. These results collectively underscore the critical role of dopant selection in tailoring the defect thermodynamics and electronic behavior of MZF. Oxygen vacancies in MZF i.e. Ov-MZF and their metal/non-metal-doped derivatives (M@Ov-MZF) induce defect states within the electronic bandgap, as evidenced by the density of states profiles and band structures in Figs S7 and S8. These defect states act as trapping centers for charge carriers (electrons and holes), thereby reducing their effective mobility. Specifically, in systems with oxygen vacancies, the formation of deep trap states leads to carrier localization, which suppresses band conduction and consequently diminishes bulk conductivity. Furthermore, in ferrites where charge transport predominantly occurs via electron hopping between mixed-valence $Fe^{2+}$ and $Fe^{3+}$ ions (e.g., in magnetite, $Fe_3O_4$) oxygen vacancies disrupt the Fe-O-Fe bridging networks. This structural perturbation breaks critical electron-hopping pathways, further degrading electrical conductivity. This dual mechanism (carrier trapping and hopping-path disruption) highlights the critical role of oxygen stoichiometry in governing the charge transport properties of MZF.

**Optical absorption**

The optical behavior of $Mn_{0.5}Zn_{0.5}Fe_{2-x}O_4$ ferrites doped with foreign elements at specific crystallographic sites (a) octahedral, (b) tetrahedral, and (c) both sites in the presence of oxygen vacancies was thoroughly explored through analysis of the complex dielectric function, as detailed in Ref. [38]. The imaginary part of the dielectric function ($\varepsilon_i(\omega)$) captures the material's light absorption capabilities, while the real part ($\varepsilon(\omega)$) is obtained via the Kramers-Kronig relations, offering insight into its dispersive response. Absorption spectra ($\alpha(\omega)$), calculated using Eq. (4) and illustrated in Fig. 4(a)–(d), span the visible wavelength range of 400–800 nm for all studied systems. These spectra exhibit clear, systematic shifts in absorption edges and peak intensities with doping content, directly tied to modifications in the electronic structure and optical bandgap. Such trends not only underscore the strong interplay between site-specific doping and optical properties, but also demonstrate the powerful tunability of these materials. This study presents optical absorption spectra analyses of MZF and Ov-MZF substituted with tetravalent ($Si^{4+}$), trivalent ($Co^{3+}$), and divalent ($Ca^{2+}$, $Mg^{2+}$, $Sn^{2+}$) ions probed via UV-Vis spectroscopy. Replacing $Fe^{3+}$ with $Si^{4+}$ creates a local charge imbalance of +1 per substitution (as $Si^{4+}$ provides one less positive charge than $Fe^{3+}$). This charge imbalance is often compensated by the formation of oxygen vacancies ($V_O$) to maintain overall charge neutrality. These oxygen vacancies ($V_O$) introduce defect states within the material's bandgap (Fig S4). These defect states increase mid-gap optical absorption, leading to a higher imaginary part of the dielectric constant (↑ ε2).

Additionally, the presence of $Fe^{2+}$ (which can form near $V_O$ or due to reduction) enables $Fe^{2+} \rightarrow Fe^{3+}$ electronic transitions, further enhancing conductivity losses and increasing ε2. Replacing $Co^{3+}$ with $Fe^{3+}$ introduces $Fe^{3+}$ ions into the crystal structure. The strong ligand field experienced by the $Co^{3+}$ ions (in their original sites) splits their d-orbitals. This d-orbital splitting creates new, distinct electronic energy levels (Fig S2 and S3). Transitions between these split d-orbitals result in new electronic absorption features. The substitution-induced modifications to the electronic structure such as bandgap narrowing (Table S1 and S2), mid-gap defect states, (Fig 2 and 3) and charge-transfer transitions were systematically. Divalent ($Ca^{2+}$, $Mg^{2+}$) and tetravalent ($Si^{4+}$) substitutions were found to modulate absorption edges in the visible-to-near-infrared range (Fig 4(a)-(d), while trivalent $Co^{3+}$ doping introduced distinct absorption features attributed to ligand-to-metal charge-transfer (LMCT) transitions. These results highlight the critical role of dopant valency and ionic radius in tailoring the optoelectronic response of MnZn-ferrites, with implications for magneto-optical devices, and energy-harvesting applications. The data reveal a substantial enhancement in optical absorption upon doping with Mg, Co, or Ca compared to the undoped MZF baseline. Among the doped samples, Mg-doped MZF exhibits the most significant enhancement, showing a broad absorption peak spanning approximately 0.2 eV to 0.8 eV. Co-doped and Ca-doped MZFs also display strong absorption features, with prominent peaks centered on 0.3 eV.

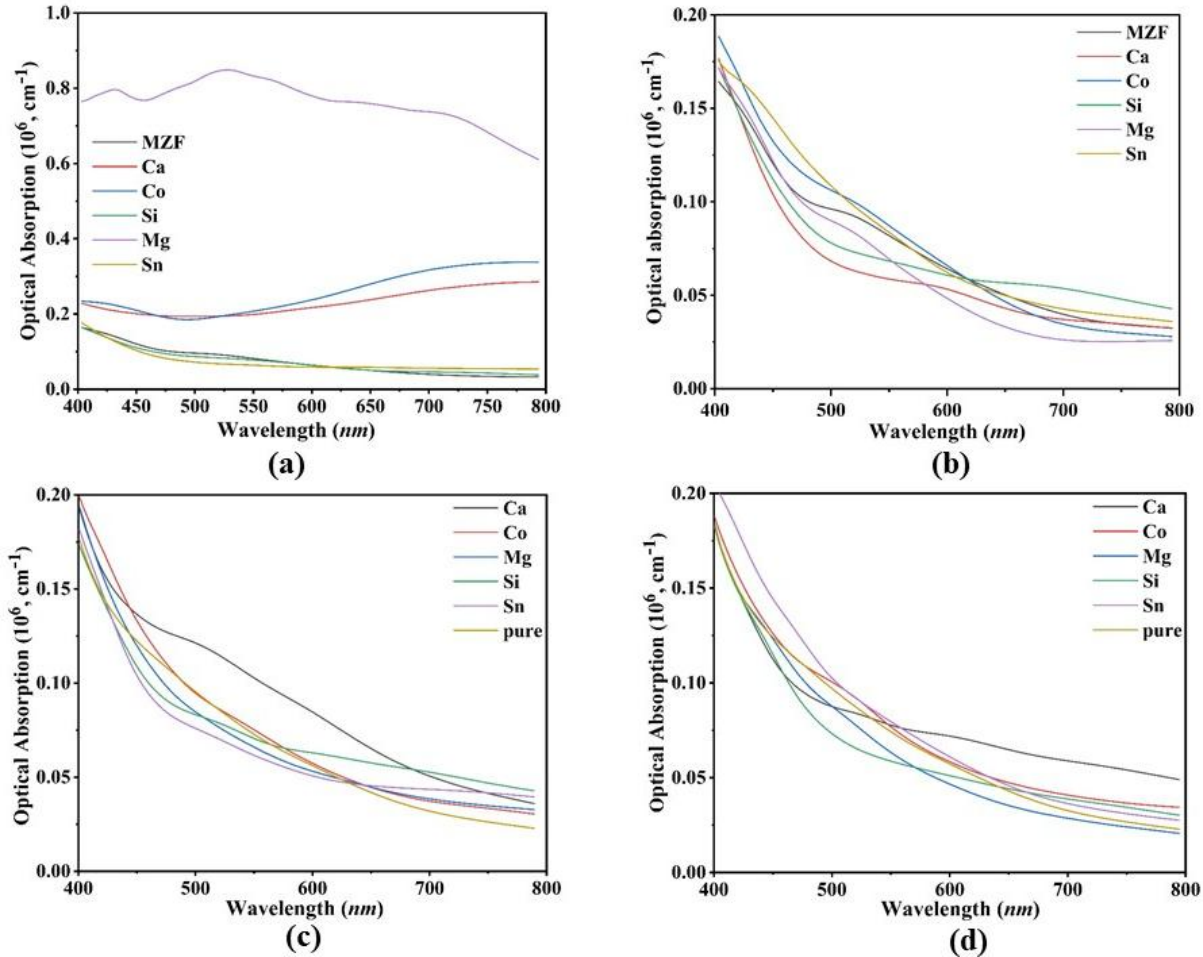

**Fig 4.** Calculated optical absorption spectra for (a) and (b) are the foreign element doped(M=Ca, Co, Si, Mg, and Sn) in Oh and Td sites of MZF; (c) and (d) are the foreign element doped (Ca, Co, Mg, Si and Sn)in Oh and Td sites of oxygen vacant $Mn_{0.5}Zn_{0.5}M_xFe_{2-x}O_{4-x}$.

For all doped variants, the absorption intensity increases sharply to a peak and then declines rapidly, tapering off to values below 0.1 eV at higher photon energies. In contrast, oxygen-deficient M@Ov-MZFs exhibit distinct absorption behavior. These samples, engineered with oxygen vacancies, show lower-intensity peaks near 0.05 eV, indicating a shift in light-matter interaction due to defect states. The pronounced absorption features across the infrared region (below 0.8 eV) underscore the strong interaction between incident light and the modified ferrite structures. Overall, the enhanced and tunable absorption characteristics driven by both dopant incorporation and defect engineering highlight the potential of doped MZFs as promising materials for photodetection and sensing applications in specific infrared spectral bands. These findings position them as strong candidates for integration into next-generation optoelectronic devices. This finely controlled modulation of optoelectronic characteristics through halide substitution positions these doped MZF as highly promising candidates for next-generation photovoltaic and light-emitting technologies.

**Magnetic anisotropic energy studies**

Magnetic anisotropic energy (MAE) plays a key role in their performance in applications like inductors, transformers, and microwave devices. Understanding and controlling MAE in soft ferrites is crucial for optimizing their magnetic properties for modern applications. Density functional theory (DFT) calculations were employed to determine the magnetic anisotropy energy (MAE) of MnZn-doped ferrites and oxygen vacancy (Ov)-doped ferrites. For each structure, the most stable geometry corresponding to the lowest MAE was identified through structural optimization under the DFT framework to obtain the minimum-energy structure. The magnetic anisotropy originates from the spin-orbit coupling of constituent elements. The magnetic anisotropy energy (MAE) is defined as the difference in the system's free energy when the spins are aligned along the z-direction compared to the xy-plane. In our study, the MAGMOM tag is used to specify the initial local magnetic moments, which are aligned along the z-direction by default, with the x and y components set to zero. The SAXIS tag defines the spin quantization axis in cartesian coordinates (u, v, w), thereby setting the direction of spin polarization. To calculate the magnetic anisotropy energy (MAE), we rotate the spin orientation along different crystallographic directions (e.g., [001], [100], [110]) using the SAXIS tag and compute the total energy for each spin orientation. Further, we determine the MAE by calculating the energy differences between these orientations.

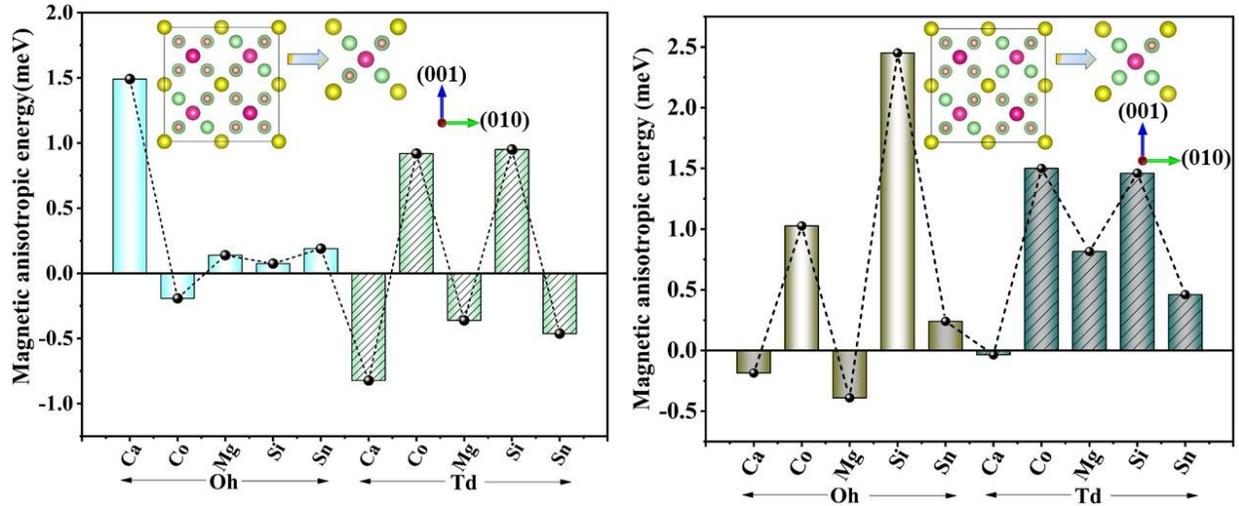

**Fig 5.** (a) Calculated magnetic anisotropy energy (MAE) for foreign element doped in MnZn-ferrite (MZF) at both octahedral and tetrahedral sites. (b) MAE for M-doped MZF at specific crystallographic sites in the presence of oxygen vacancies (Ov-MZF).

DFT+U calculations reveal that both pristine $Mn_{0.5}Zn_{0.5}Fe_2O_4$ (MZF) and its oxygen-deficient counterpart exhibit negligible magnetic anisotropy energy (MAE), with values effectively converging to zero. In contrast, doped MZF systems both stoichiometric and oxygen-vacancy-containing ($Mn_{0.5}Zn_{0.5}Fe_{2-x}O_{4-x}$) exhibit finite MAE values that vary in sign and magnitude depending on the dopant species and local structural configuration. The magnetic anisotropy in soft ferrites such as MnZn-ferrite is highly sensitive to the distribution of cations across the tetrahedral (A) and octahedral (B) lattice sites. Substitutional doping alters this cation distribution, which in turn influences the magnetocrystalline anisotropy constant ($K_1$). By carefully tuning the site-specific dopant incorporation, $K_1$ can be reduced, thereby minimizing the MAE a critical requirement for lowering magnetic losses in high-frequency applications such as transformers and inductors.

In doped MZF systems with balanced tetrahedral and octahedral coordination environments, the computed MAE values span a range from approximately +1.8 meV to –0.72 meV (Fig. 5a). This variation reflects competing magnetocrystalline anisotropy contributions originating from different sublattices. Negative MAE values correspond to energetically favored easy-axis magnetization directions, while positive values indicate harder axes that resist magnetization. The presence of both signs underscores the complex interplay between dopant type, site occupancy, and local symmetry distortions, highlighting the tunable nature of MAE in engineered MnZn ferrites. In O-vacant MZF-doped ferrites, the MAE range widened to <2.4 meV to <-0.45 meV, (Fig 5(b)) reflecting stronger anisotropy fluctuations due to oxygen deficiency. The higher positive MAE peak (2.4 meV vs. 1.8 meV) suggests enhanced anisotropy along certain axes, likely due to disrupted superexchange interactions from missing oxygen ligands. The less negative MAE (-0.45 meV vs. -0.72 meV) indicates a weaker easy-axis preference, possibly due to altered $Fe^{2+}/Fe^{3+}$ valence states and localized spin reordering. The narrower MAE spread in octahedral-tetrahedral ferrites suggests more uniform anisotropy, favoring lower coercivity ($H_c$) and reduced hysteresis losses. The broader MAE distribution in O-vacant ferrites implies higher magnetic disorder, potentially increasing eddy current losses at high frequencies. Optimal doping should minimize oxygen vacancies to stabilize low MAE values for high-performance soft ferrite applications. First-principles studies to correlate MAE trends with specific cation-disorder configurations. Experimental validation via torque magnetometry or FMR need to be confirm theoretical predictions. Defect engineering to suppress O-vacancy-induced anisotropy while maintaining desired electrical resistivity.

**Electrical properties**

MnZn-ferrites are soft ferrites widely used in power electronics due to their high permeability and low core losses. Their electrical conductivity is a critical property affecting eddy current losses and overall performance. Enhancing the thermal properties of soft magnetic materials is crucial for advancing electronic devices toward higher frequencies, improved efficiency, and greater miniaturization. The thermoelectric transport properties, including the electrical conductivity ($\sigma$), Seebeck coefficient (S) were calculated using first-principles density functional theory (DFT) coupled with Boltzmann transport theory as implemented in the BoltzTraP code. This approach provides a rigorous framework for investigating charge carrier transport under temperature gradients and electric fields. The computed electrical conductivity ($\sigma$) and Seebeck coefficient (S) of foreign-element-doped MnZn-ferrites (MZF) at 300 K are presented in Fig 6(a) and (b), illustrating dopant occupation at both octahedral and tetrahedral sites, with corresponding oxygen vacancy configurations detailed in Fig 6(c) and 6(d). Analysis of Figs 6(a) and 6(b) reveals that foreign-element doping at both crystallographic sites reduces electrical conductivity relative to undoped MZF, while simultaneously enhancing the Seebeck coefficient (S). This inverse relationship between $\sigma$ and S suggests a doping-induced trade-off, likely arising from increased carrier scattering and modified band structure effects. Computed properties at 300 K: Pure MZF exhibits an electrical conductivity of $1.961 \times 10^{23}$ $\Omega$ m$^{-1}$ K$^{-1}$ and a Seebeck coefficient of mVK-1. Doping with Ca increases the Seebeck coefficient to 12.667 µV K$^{-1}$ (similar for both sites). The electrical conductivity depends on the Ca occupation site: $1.267 \times 10^{23}$ $\Omega$ m$^{-1}$ K$^{-1}$ (octahedral, Ca@Oh) and $1.373 \times 10^{23}$ $1.267 \times 10^{23}$ $\Omega$ m$^{-1}$ K$^{-1}$ (tetrahedral, Ca@Td). These results underscore the critical interplay between dopant site occupancy, oxygen vacancy architecture, and thermoelectric performance in MZF systems.

Notably, the oxygen vacancy distributions mapped in Figs 6(c) and (d) correlate with these trends, implying that vacancy-mediated carrier localization and altered $Fe^{2+}/Fe^{3+}$ hopping pathways contribute to the suppression of conductivity. The enhanced Seebeck coefficient, conversely, may originate from energy-dependent carrier filtering at defect states or a shift in the Fermi level due to dopant-induced charge compensation. Interestingly, in the oxygen-vacant MZF (Ov-MZF) system, doping with Co or Mg at octahedral sites slightly enhances electrical conductivity compared to undoped Ov-MZF structure, Co-doped (Co@Oh): $1.306 \times 10^{23}$ $\Omega$ $m^{-1}$ $K^{-1}$, Mg-doped (Mg@Oh): $1.292 \times 10^{23}$ $\Omega$ $m^{-1}$ $K^{-1}$ and Undoped Ov-MZF: $1.231 \times 10^{23}$ $\Omega$ $m^{-1}$ $K^{-1}$.

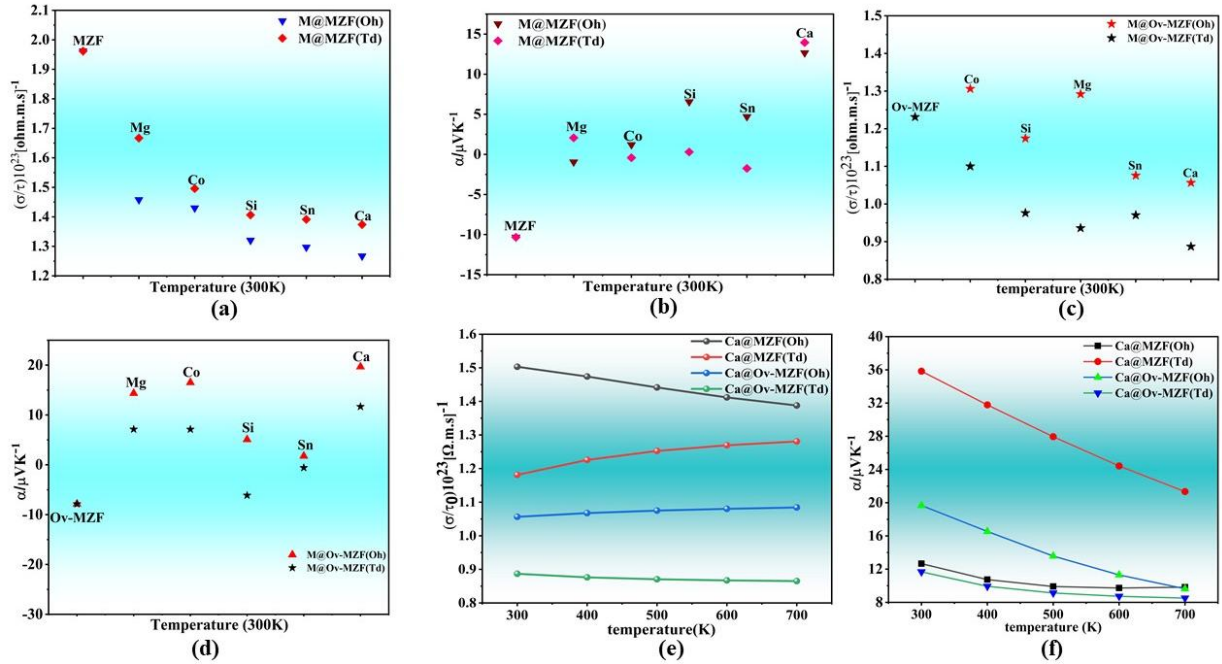

**Fig 6.** Electrical transport characteristics of doped-MZFs: (a, b) Electrical conductivity and Seebeck coefficient at 300 K for MZF doped with various foreign elements, considering both octahedral and tetrahedral site substitution. (c, d) Electrical conductivity and Seebeck coefficient at 300 K for doped Ov-MZF, with dopants occupying octahedral and tetrahedral sites. (e, f) Temperature-dependent electrical conductivity and Seebeck coefficient for Ca-doped MZF (Ca@MZF) and oxygen-vacant Ca-doped MZF (Ov-Ca@MZF), comparing site-specific dopant effects.

Doping of foreign elements (Ca, Si, Mg, Co, Sn) at tetrahedral and octahedral sites in $Mn_{0.5}Zn_{0.5}Fe_{2-x}O_4$ induces significant lattice distortion and substantially alters electrical transport properties, including electrical conductivity and Seebeck coefficients. These changes correlate with dopant-driven structural modifications: lattice expansion (Ca, Sn, Mg, Co) or contraction (Si) directly impact carrier mobility and concentration, as evidenced by temperature-dependent electrical property measurements (Fig 6 (a)-(f). Both Ca and Sn doping at tetrahedral or octahedral sites increase electrical resistivity relative to pure MZF and oxygen-vacant MZF (Ov-MZF), as evidenced by reduced electrical conductivity and altered Seebeck coefficients. Furthermore, we analyze the temperature-dependent (300-800K) electrical conductivity and altered Seebeck coefficients for the high-resistivity Ca-doped MZF system. Fig 6(e) presents the computed temperature-dependent electrical conductivity for Cu-doped MZF (Cu@MZF) and oxygen-vacant Cu@MZF (Ov-Cu@MZF), comparing octahedral (Oh) versus tetrahedral (Td) site doping effects. The conductivity exhibits minimal variation across the 300-800 K temperature range. Fig 6(f) shows the corresponding Seebeck coefficients, which decrease monotonically with increasing temperature.

**Conclusions**

This study presents a comprehensive first-principles investigation into the effects of multivalent ion substitution and oxygen vacancy engineering on the structural, electronic, magnetic, and thermoelectric properties of MnZn ferrites ($Mn_{0.5}Zn_{0.5}Fe_2O_4$). Using density functional theory (DFT), we systematically explore how the incorporation of $Ca^{2+}$, $Si^{4+}$, $Mg^{2+}$, $Co^{3+}$, and $Sn^{4+}$ ions, as well as oxygen vacancies (Ov), influences key material characteristics across different crystallographic sites. Formation energy calculations reveal that several substituted systems, particularly those doped with Ca, Si, and Mg, exhibit negative formation energies (ranging from ~ –1.03 eV to –3.24 eV), indicating favorable thermodynamic stability and potential experimental reliability. Lattice distortion analyses show that dopant-induced strain is governed by ionic radius, valence mismatch, and site preference. Substituents with ionic radii similar to $Fe^{3+}$, such as $Co^{3+}$

(0.72 Å), result in minimal lattice distortion ($\Delta a < 0.5\%$), whereas larger ions like $Ca^{2+}$ (1.00 Å) and $Sn^{4+}$ (0.55 Å) induce notable lattice expansion ($\Delta a \approx 1.8$–$3.1\%$). Conversely, the undersized $Si^{4+}$ ion causes a unique lattice contraction ($\Delta a \approx -0.7\%$). Electronic structure calculations confirm that all substituted MZF systems maintain finite band gaps, preserving their semiconducting nature. However, oxygen-vacancy-containing variants (Ov-MZF) exhibit narrower band gaps due to the emergence of defect-induced mid-gap states, which significantly influence carrier transport. These defect states contribute to an increase in the Seebeck coefficient and a reduction in electrical conductivity, reflecting a doping-induced trade-off that can be leveraged for thermoelectric optimization.

Magnetic anisotropy energy (MAE) analyses indicate that targeted engineering of octahedral and tetrahedral site occupancy leads to more uniform anisotropy distributions, which is beneficial for low-coercivity, soft magnetic behavior essential for high-frequency applications. Among all dopants examined, Ca and Sn substitutions yield the most promising thermoelectric enhancements over pristine MZF, characterized by reduced electrical conductivity and elevated Seebeck coefficients. The findings establish a predictive framework linking defect chemistry and site-selective doping to property tuning in spinel ferrites. These insights offer strategic guidance for the design of next-generation ferrite materials with optimized performance for energy conversion, thermoelectric applications, and high-frequency magnetic devices. Future efforts should prioritize systematic experimental validation and magneto-transport measurements to corroborate the theoretical predictions presented in this study.